# Recent Photometric Monitoring of KIC 8462852, the Detection of a Potential Repeat of the *Kepler* Day 1540 Dip and a Plausible Model

R. Bourne[1], B. L. Gary[2], and A. Plakhov[3]

[1]*East Cannington, Perth, Western Australia 6107*
[2]*Hereford Arizona Observatory, Hereford, AZ, 85615, USA*
[3]*Krasnogorsk, Moscow, Russia 143401*

**ABSTRACT**

This paper presents V- and g'-band observations of the F2V star KIC 8462852, which exhibited enigmatic fade patterns in *Kepler* mission data. We introduce a transit simulation model for interpretation of these fades, and use it to interpret an August 2017 dip as a repeat of the *Kepler* day 1540 dip (D1540). We suggest the August 2017 and D1540 dips may be caused by a brown dwarf and an associated ring system in a 1601-day elliptical orbit. Transiting icy moons of the proposed brown dwarf, sublimating near periapsis like comets, could provide an explanation for the significant dips observed by *Kepler*, as well as the recent May to October 2017 dips and the long term variation in flux detected by Simon at al. (2017). Whereas the presence of such a ring structure is attractive for its ability to explain short term fade events, we do not address how such a ring system can be created and maintained. This speculation predicted a brightening of ~1 % that occurred during October 2017. In addition, this scenario predicts that a set of dimming events, similar to those in 2013 (*Kepler*) and in 2017 (reported here), can be expected to repeat during October 2021 to January 2022 and a repeat of D1540 should occur on 27 December 2021.

E-mail: blgary@umich.edu

## 1. INTRODUCTION

The main-sequence F2V star KIC 8462852, hereafter KIC8462, experienced several unusual dimming events during the 4 years of observation by *Kepler* (Boyajian et al. 2016). The unique dimming events of a mature main sequence star lacking infrared (IR) excess are unprecedented and have led to numerous model speculations. A variety of possible explanations for the anomalous light curve were examined by Boyajian et al. (2016), where it was tentatively suggested that the most likely cause for the fade behavior was a swarm of comets. KIC8462 is estimated to have a stellar mass of 1.43 $M_s$, and a radius of 1.58 $R_s$ (Boyajian et al. 2016), which we adopt for simulations undertaken in this paper.

Current explanations fall short in providing an adequate simulation of the shape and duration of the major dips in the *Kepler* data together with a justification of the 3 year *Kepler* dimming trend, followed by a 2.4 % fade, reported by Montet and Simon (2016), hereafter MS16. The long term variability in ASAS and ASAS-SN data found by Simon et al. (2017) also remains unexplained.

Meng et al. (2017) suggest that KIC8462 may have faded during a 1-year interval based on observations in UV-, B-, V-, Rc-bands and the Spitzer spacecraft channels at 3.6 and 4.5 micron. No significant IR excess was identified from the Spitzer observations.

Sacco et al. (2017) interpreted the 2017 dips as a repeat of the 2013 *Kepler* dips with a period of 1574 days by correlating the 2013 and 2017 light curves. However, objects with identical orbital periods are only stable at Lagrange regions, not in close proximity to each other as proposed by Sacco et al.

Kenworthy and Mamajek (2105) developed a model for fitting the transit observations of SWASP J1407. It uses flux gradients and transit depth to derive parameters of individual rings, surrounding a massive object, and succeeded in producing a plausible explanation for the several weeks of brightness variations. Aizawa et al. (2017) developed a different model, involving a precise integration scheme to compute transit light curves by a ringed planet.

In Section 2 we extend the Aizawa et al. (2017) precise integration technique of simulating photometry light curves by planets with ring systems to develop a light curve simulation model: Simulated Photometry of Transits (SPOT). It is based on 2D rendered objects occluding a 2D rendered limb darkened stellar disk.

The Hereford Arizona Observatory (hereafter HAO) commenced observing KIC8462 from October 2015 to November 2017. Section 3 describes V- and g'-band observations from May to November 2017, when a series of complex dimming events were observed.

In Section 4 modeling is undertaken for the D1540 dip using SPOT to identify and estimate the key parameters of a potential brown dwarf and ring system. It was found that a good fit was obtained for a ring system that extended beyond the Roche limit; any such ring system would therefore be transient in nature.

Section 5 presents a comparison of the HAO August observations with the SPOT model for D1540. Excellent agreement was found between the shapes and depths of the two events, and we interpret these as repeat transits of the same object.

In Section 6 we place some initial constraints on the orbit of the proposed brown dwarf based on transit velocities, orbital period and the long term flux variations.

The authors want to emphasize that this paper's purpose is not to promote a ring model to the exclusion of other models. It is limited to a demonstration that two of the most prominent dips, observed about 1600 days apart, can be easily accounted for with a simple and natural model (rings). By implication, other dips may also be explainable with such a model. We will not argue the case for the likelihood that ring systems, sublimating moon comas and dust clouds are the sources for these unusual transits because calculations of mass loss and escape velocities for populating these transiting structures is beyond the scope of this paper. Finally, in a future paper we intend to present a possible configuration of the overall system in a way that provides a potential explanation for the 3-year 2 % fades at intervals of 1600 days.

Notes:

*Kepler* light curves presented here are constructed from either the "Normalised Flux" data at http://www.wheresthefiux.com/public or from the Mikulski Archive for Space Telescopes website http://archive.stsci.edu/kepler/data_search/search.php.

HAO observations are available at:

http://www.brucegary.net/KIC846/#2017.06.15_V,
http://www.brucegary.net/ts/,
http://www.brucegary.net/ts3/
http://www.brucegary.net/ts4/
http://www.brucegary.net/ts5/

## 2. SIMULATED PHOTOMETRY OF TRANSITS (SPOT) MODEL

SPOT uses a ring transit simulation technique similar to the one developed by Aizawa et al. (2017), precise integration, and consists of a 2D rendered limb-darkened stellar disk occluded by a 2D image of a planet and rings, pre-rendered based on input parameters, to generate a measurement of flux relative to the overall flux of the limb-darkened stellar disk. Using 2D images to represent the limb-darkened stellar disk and the planet and ring system saves computing resources in comparison with 3D models, facilitating extensive modeling of a range of ring configurations including those beyond the Roche limit out to the Hill Sphere.

The SPOT model uses the 'Quadratic Limb Darkening Law' from Sing (2009)

$$I(\mu) / I(1) = 1 - a \times (1-\mu) - b \times (1-\mu)^2 \quad (1)$$

Where $I(1)$ is the intensity at the center of the stellar disk, $\mu = \cos(\theta)$ or $\sqrt{1-r^2/R^2_*}$, and *a* and *b* are the Limb Darkening Coefficients (LDCs). The LDCs for KIC8462 used in this paper, **a = 0.2672** and **b = 0.3267**, are taken from Sing's *Kepler* LDCs for a star with $T_{eff}$ = 6750.

The simulation consists of two independent phases: a rendering phase and a simulation phase. In the rendering phase, a textual configuration of an occluder is converted into a large 2D array representing a plane perpendicular to the line of sight. Every value in this array (further referred to as a "pixel") defines a percentage of luminosity that will be blocked by this occluder in a corresponding square part of this plane. This can be less than 100% even for a completely opaque occluder (for example, a planet), if a pixel lies on the occluder's edge.

Physical dimensions of a pixel can be set in a configuration file, in a typical simulation they are between 1,000 x 1,000 km to 5,000 x 5,000 km. Dimensions for a large occluder can be thousands of pixels on each axis and the number of nonzero pixels can be in the millions.

The main assumptions of the simulation are that:
- the rendered array remains constant during the transit;
- an occluder's transit velocity across the stellar disk is constant; and

- any changes in the configuration of an occluder's components (impact parameter, tilt, obliquity, opacity etc) during the transit can be neglected.

Clumps of dust/ice external to a ring and spokes/holes in rings that rotate through the ring during a transit are not accounted for in this simulation and could account for variation between the model and an actual light curve.

A second 2D array is prepared for the star, where every pixel is assigned a luminosity up to 1 depending on its location in the stellar disk in accordance with the quadratic limb darkening formula. The next component of the simulation superimposes the two 2D arrays with a varying shift according to the transit speed. For every stellar pixel its limb darkened luminosity value is multiplied by the occluding pixel's transparency.

The simulation model is contained in a single executable file and input parameters are specified in a text configuration file. The model generates a video display of the planet and rings occluding the star, a 2D rendered image of the occluder and a text file of the calculated flux data points.

Pixel size (number of pixels equal to the stellar radius) and frame rate (number of pixels between data points) are the key user specified parameters that determine the overall accuracy of the simulation and run times. User specified input parameters for the star, planet and ring system generate a simulated light curve (flux data-points) and a video preview of the simulated transit.

Input parameters for the stellar component of the model include:
- stellar radius,
- impact parameter, and
- limb darkening.

Input parameters for the ring and planetary component of the model include:

- planetary radius
- outer radius and ring width,
- eccentricity,
- apparent tilt,
- apparent obliquity, and
- opacity.

The model allows for up to nine rings to be modeled concurrently and can be adapted to simulate other complex transiting objects such as dust clouds. Opacity is specified at face-on orientation and can vary from 0 (clear) to 1 (fully opaque). The opacity is automatically adjusted according to the apparent obliquity.

Planetary transits have distinct U shaped light curves that vary in shape depending on the impact parameter. The impact of planetary transits on stellar flux is limited due to the size limit of planet radii of approximately 70,000 km, Chen and Kipping (2016), excluding inflated hot Jupiters and planets in the process of formation.

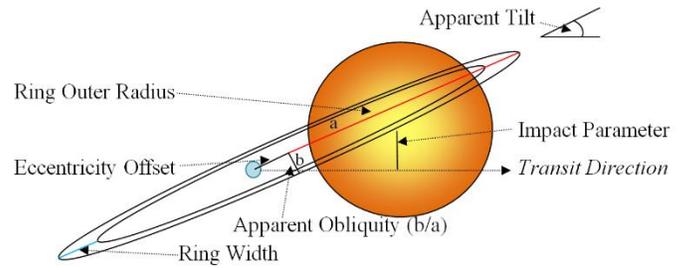

**Figure 2.1:** SPOT input parameters

Unlike planetary transits, ring transit light curves can have a wide range of shapes and distinctive features depending on orbital radius, tilt, obliquity, width and opacity. The depth of ring transits is only limited by the size and opacity of rings. While long term stable ring structures are restricted to the Roche limit of 2.45 planetary radii (Schlichting and Chang, 2011), transient rings like Saturn's Pheobe ring are limited to approximately 0.46 times the Hill Sphere for pro-grade orbits and 0.75 times the Hill Sphere for retro-grade orbits (Rieder and Kenworthy, 2016).

A detailed description of the SPOT model is proposed as a separate paper including the public release of the program used in this paper to simulate ring transits. While the initial release may be restricted to ring and planetary simulations, enhancements to SPOT to include simulation modules for comas and dust cloud transits, secondary transits and automated light curve matching are under development.

## 3. 2017 OBSERVATIONS BY THE HEREFORD ARIZONA OBSERVATORY

The Hereford Arizona Observatory (HAO) is located in Hereford, Arizona, (https://en.wikipedia.org/wiki/Hereford_Arizona_Observatory). A 14-inch Meade LX200GPS telescope in an ExploraDome was used for the present observations. MaxIm DL 6.2 software (MDL) drives an after-market SciTech telescope mount controller card via 100-foot buried cables. MDL also controls the dome, a wireless MicroTouch focuser and a Santa Barbara Instrument Group ST-10XME CCD camera. The field-of-view is 15 x 10 'arc, and image scale is 0.83 "arc/pixel when binned 2x2. All exposure times were 30 seconds. Focus for sharp images is maintained automatically (i.e., the defocus observing strategy would be dangerous in the KIC8462 crowded star field).

Observing session images are calibrated using master bias, dark and flat images. After star-alignment, photometry measurements are made using a fixed set of aperture radii: 7, 10 and 12 pixels for signal aperture radius, gap width and sky background annulus width. The signal photometry size is intentionally not changed when "atmospheric seeing" changes the "point spread function" (PSF) size because KIC8462 is in a crowded star field and it is important to prevent the incursion of nearby star PSFs into the signal aperture.

Observations were made using a V filter from 2015 October 16 to 2017 September 20, after which a g'-band filter was used. As many as 34 nearby stars were evaluated for use in calibration, but only 18 were stable at the milli-magnitude (mmag) level, so that's the subset of nearby stars that have ended-up being used. An analysis of self-consistency during several months of observations was used to assign a weight to each reference star when performing a "star color sensitivity" calibration fit.

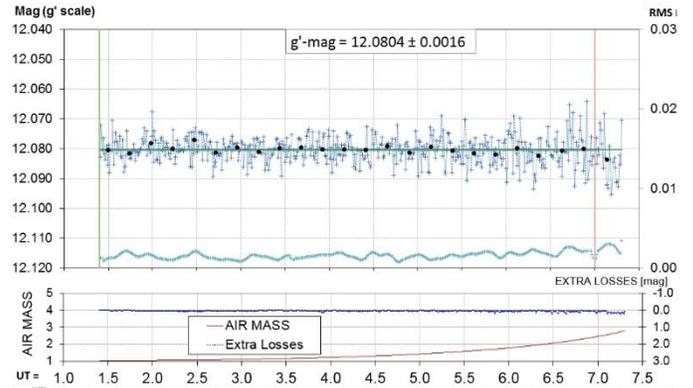

**Figure 3.2.** Light curve for an observing session. The upper panel is a LC for g'-magnitude readings for each image, with symbols for 21-point averages. The noise level for KIC8462 is estimated from a magnitude model for the internal-consistency noise level of 3 nearby reference stars. The lower panels shows "extra losses" (due to clouds, for example).

Precision (day-to-day repeatability) was typically 1.2 mmag for V-band, and 0.7 mmag for g'-band. Figure 3.3 shows the V- and g'-band measurements for 6 months in 2017 when dips were observed.

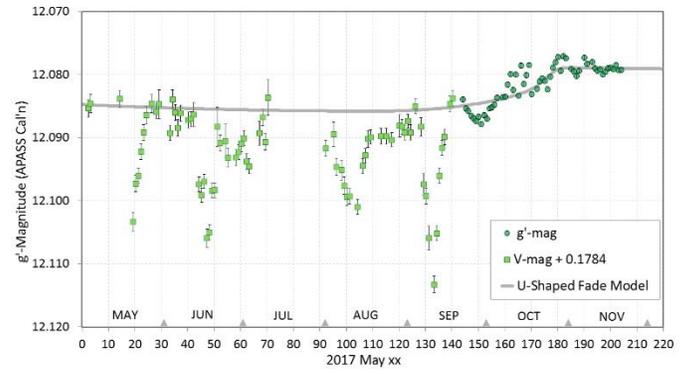

**Figure 3.3.** Measured g'-magnitude (and V-magnitude converted to g'-magnitude) for 6-months in 2017. The "U-Shaped Fade Model" is a long-term variation model (gray trace) which is used as a reference for determining dip depth.

The out-of-transit (OOT) brightness model, referred to in Fig. 3.3 as the "U-Shaped Fade Model", has a smooth variation that "skims the tops" of the 6.5 months of measurements shown in the figure. The "U-Shaped Fade Model" is shown for a longer span of dates in Fig. 3.4.

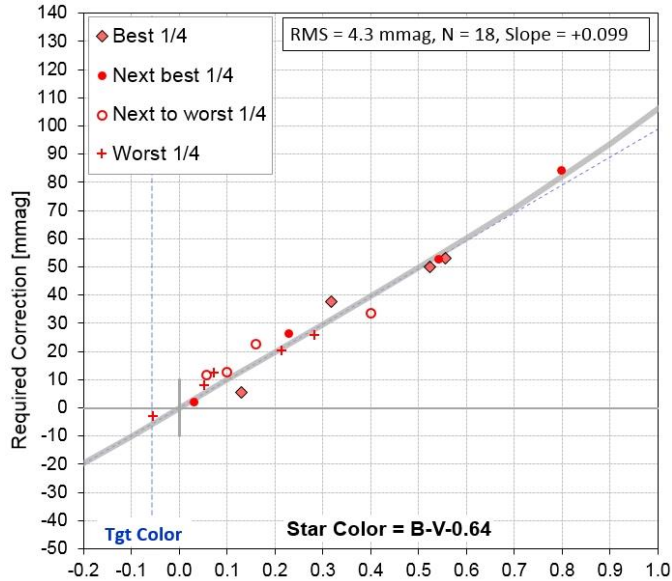

**Figure 3.1.** Instrumental magnitude vs. star color, used for calibrating an observing session's magnitude offset.

Figure 3.1 presents a typical observing session plot of instrumental magnitude vs. star color for 18 nearby calibration stars, with a model fit for estimating magnitude offset for the target star (KIC8462) at its adopted star color. Stars are weighted using 1/RMS_Precision in order to emphasize the influence of stars that are well-behaved during a several-month analysis interval.

A typical observing session light curve (LC) is shown in Fig. 3.2. It is constructed in an Excel spreadsheet developed during the past 15 years specifically for analysis of exoplanet transits.

The U-Shaped Fade Model in Fig. 3.4 is a mathematical construction meant to represent OOT measurements; it is not based on a physical model for whatever is producing the long-term variations. The model consists of a linear fading trend between U-shaped fade events (inspired by the MS16 *Kepler* full frame image analysis), and cosine functions (raised to 0.5 power) for the first and second segments of the U-shaped fade feature. A distinction is made between measurements that are determined to be unaffected by fading due to the transit of something (such as a dust cloud) and measurements determined to be affected by such transits. In Fig. 3.4 measurements identified as belonging to the two states are referred to as OOT and dip.

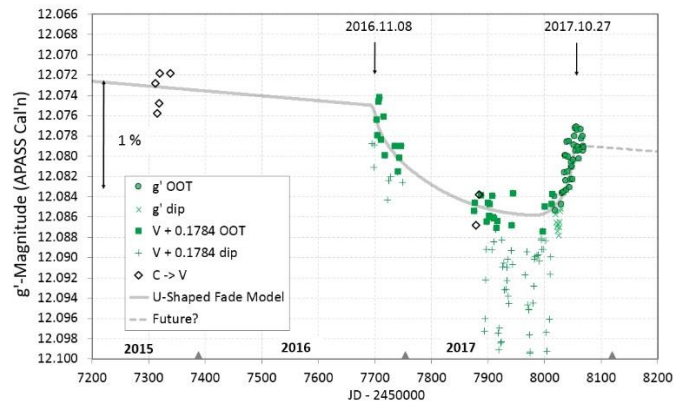

**Figure 3.4:** A 2-year light curve of HAO observations where a distinction is made between OOT measurements (filled symbols for V- and g'-band and open diamonds for unfiltered observations) and dip measurements (green + and x symbols).

Based on MS16 and the 2017 "U-Shaped Fade Model", KIC8462 appears to undergo a 1-year fade approximately every 1600 days. It is also noteworthy that most dips occur during the last 6 months of these big fade events. A search for a repeat of specific *Kepler* dips in the HAO data is discussed in Section 5, where the August dip is interpreted as a repeat of D1540 with a period of 1601 days.

Simon et al. (2017) identified a brightening after the *Kepler* observations using ASAS and ASAS-SN data. Another period of brightening just before the commencement of Kepler observations is seen in Fig 4 of the Simon et al. (2017) paper. Together, these two long timescale brightening patterns lend support to our suggestion of an approximate 1600 day periodicity.

A brightening feature would not have been seen in the *Kepler* data because of the *Kepler* hardware failure that ended the main mission, which is shortly before our 1601-day repeat interval would predict a brightening. Therefore, its absence in the MS16 full frame image analysis of *Kepler* data is compatible with the "U-Shaped Fade Model" that repeats at ~1600-day intervals.

There is a small difference between the length of the U-shaped fade feature in the current year LC, which is 1.0 ± 0.2 year, and the length of a fade event that is derived for the *Kepler* observing interval obtained by folding the *Kepler* LC with a 1601-day period. The *Kepler* data fade length is estimated to be 1.7 ± 0.3 year. The depths also differ somewhat between *Kepler* (2.4%) and HAO observations (1.1 %).

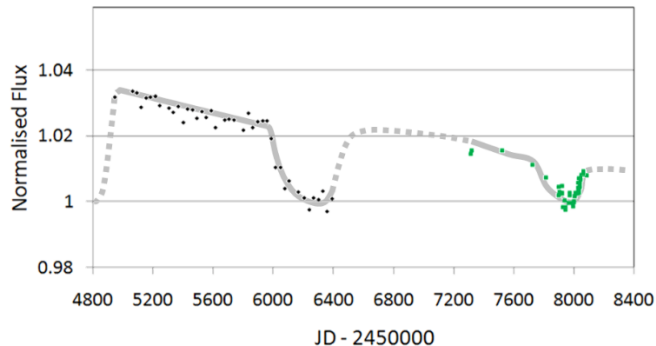

**Figure 3.5:** Comparison of Kepler mission normalized flux (MS16), black diamonds, with HAO normalized flux, green squares, showing a 1601-day repeat of the proposed "U-Shaped Fade Model".

Figure 3.4 shows that the current U-shaped egress has not reached the same brightness level that was present before ingress. The level of the dotted trace following egress is based on several measurements with the same magnitude, and the slope of this trace is arbitrarily set to have a value similar to what was present before ingress.

The same post-egress behavior is shown in Fig. 3.5. In creating Fig. 3.5 we made use of the freedom to adopt as a free parameter the magnitude offset between the *Kepler* and HAO data sets. We arbitrarily set the U-shaped minima to be the same. After adopting this offset it became apparent that the *Kepler* data pre-ingress fade rate of ~ 0.4 %/year could be extended to the time of HAO data. It's as if there is something fundamental and stable about the brightness level of the U-shaped minimum brightness, and two components of variation are superimposed upon this brightness: 1) a slowly varying longer-timescale variation, and 2) a U-shaped function that multiplies with the first component. A physical mechanism for this behavior will be discussed in a future paper.

We are not proposing an explanation for the U-shape fade feature. In fact, we are exploring an explanation that doesn't involve blocking of light, but instead relies upon the reflection of starlight by large particles that undergo changes in illumination direction during the course of an orbit. The purpose for calling attention to this fade feature in the present paper is that it supports the case for a 1601-day orbit and the same structures being responsible for both the U-shaped fade and the repeat of the D1540 dip.

## 4. KEPLER DAY 1540 DIP – A POTENTIAL BROWN DWARF AND COMPLEX RING SYSTEM

The light curve shape for the D1540 event has a prominent deep and rapid central dip and long and complex ingress and egress. The SPOT model shows this dimming event can be simulated by a brown dwarf with 9 rings extending out to approximately 0.1 AU, tilted at 0.3 degrees.

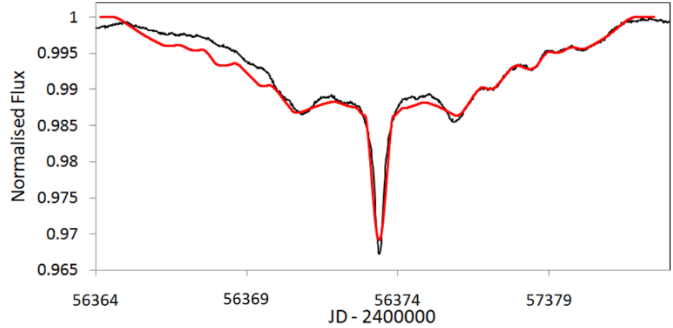

**Figure 4.1:** "Normalized flux" light curve for the D1540 dip, showing *Kepler* data (black trace) and a SPOT model fit (red trace).

Whereas the egress "ripple" pattern with 5 minima would imply that a 5-ring model should provide an adequate fit to the measured light curve, an extra ring is required to produce the "flat" section adjacent to the central dip. The depth of the central dip is considerably greater than what could be produced by a planet or BD and an additional three rings are needed to reproduce the shape and depth of this central feature. The ring modeling was undertaken in 2016 and was not fully optimized due to the inadequate fit, in terms of timing and depth, to the ingress curve.

While the match for the egress in this simulation is reasonable the ingress is not as well matched. The major contributing factor appears to be reduced opacity and/or width of the leading edge of the outer ring, resulting in an increase in flux of almost 0.2%. The trailing edge of the outer ring results in a reduction in flux of over 0.4%. This asymmetry in opacity and/or width of the outer ring could be produced by a transient ring in the process of accruing material. This asymmetry of opacity in the outer ring does not appear to be present in the August 2017 transit as discussed in Section 5.

The proposed impact parameter of 0.99 (1.09 million km), almost the full radius of the star, results in only half the ring system occluding the star. It is difficult to fully constrain the radius of the planet due to the high opacity of the inner rings. A planet radius of 70,000 km has been used in the SPOT model since this is the approximate size of all objects with masses in the range of ~ 13 to 80 times Jupiter's mass (i.e., all brown dwarfs).

The SPOT model provides an estimated transit velocity of ~ 28 km s$^{-1}$. It should be noted there is asymmetry in the timing of the ring transits that might be due to a variation in the transit velocity.

It is estimated that the actual transit velocity may have increased from approximately 27.5 km s$^{-1}$ to 28.5 km s$^{-1}$ over the course of the transit. This is in agreement with the potential 1601-day eccentric orbit (discussed in Section 6) as the proposed brown dwarf approaches periapsis.

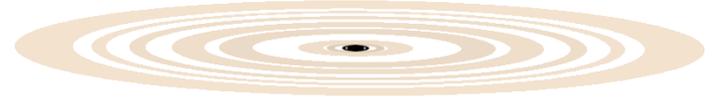

**Figure 4.2:** 2D rendered simulation model of the proposed D1540 brown dwarf ring system.

| Ring | 1 | 2 | 3 | 4 | 5 |
|---|---|---|---|---|---|
| Outer Radius (AU) | 0.002 | 0.003 | 0.007 | 0.016 | 0.039 |
| Ring Width (AU) | 0.001 | 0.001 | 0.002 | 0.007 | 0.009 |
| Opacity at 0° | > 0.03 | > 0.03 | 0.015 | 0.0035 | 0.0055 |
| Ring | 6 | 7 | 8 | 9 | |
| Outer Radius (AU) | 0.052 | 0.064 | 0.074 | 0.098 | |
| Ring Width (AU) | 0.009 | 0.004 | 0.004 | 0.019 | |
| Opacity at 0° | 0.0035 | 0.004 | 0.0018 | 0.0012 | |

**Table 4.1:** Ring Parameters of the D1540 brown dwarf ring system. Opacity input in SPOT is at 0° (face-on) and automatically converted to transit opacity depending on obliquity, this eliminates the need to manually recalculate the opacity as variations are made to the obliquity parameter during the modeling task.

The three inner rings of the proposed ring system have high opacity while the remaining rings all have low opacities. Several wide "gaps" between rings are identified by the SPOT simulation. It is unclear if these gaps are due to small orbiting planets or are the result of harmonic influences of large outer planets of the brown dwarf that we propose in a paper currently in preparation.

If the BD mass is 0.07 times solar, then the proposed ring system would extend past the Roche limit, beyond which dust forms into moons. The proposed brown dwarf rings could therefore be transient in nature and may have recently accumulated dust. If the source for the dust in these proposed transient rings is due to sublimation of icy satellites of the brown dwarf, the drown dwarf system may have migrated recently from beyond the 'snow line.'

The proposed transient ring structure at KIC8462, filling a large portion of the Hill sphere, is not the first large transient exo-ring structure to be suggested. Kenworthy and Mamajek (2015) have modeled the significant dimming in the flux of the young stellar object SWASP J1407 as an extensive ring system surrounding a potential planet J1407b. While the KIC8462 and J1407 systems may share some similarities, in that both ring systems occupy significant portions of their respective Hill spheres, the proposed J1407b ring system is considerably larger than the proposed D1540 ring system, with an outer radius of 0.6 AU compared to 0.2 AU, and a drop in flux greater than 90% compared to around 3.3% for D1540. The proposed J1407b ring system is also considerably more complex, with at least 35 distinct rings of various sizes and opacities, likely orbiting at a greater distance from the parent star and possessing a significantly larger tilt and obliquity.

## 5. AUGUST 2017 DIP – A POTENTIAL REPEAT OF THE D1540 DIP

The dimming event in August 2017 lasted over two weeks and saw a long complex ingress followed by a rapid central dip and a long complex egress. Our plot of this HAO data in Fig. 5.1 uses an adopted out-of-transit g'-magnitude of 12.086, as specified by the U-shaped fade model (c.f., Fig. 3.3).

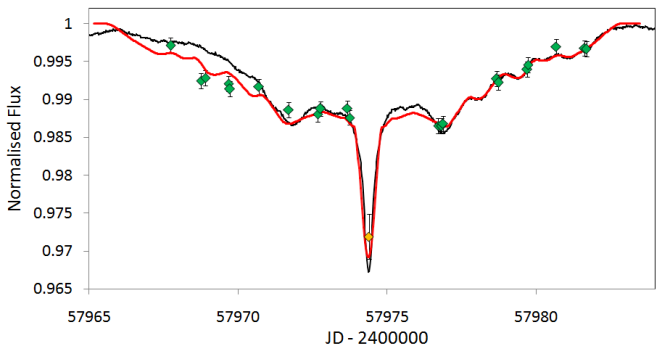

**Figure 5.1:** Normalized flux light curve for the August 2017 dip, showing *Kepler* data (black trace) shifted forward in time by 1601 days, HAO data from August, 2017 (green symbols) and a SPOT model fit (red trace) derived before the 2017 HAO observations. The 09 August 2017 (JD 2457974) observation (Boyajian, private communication, orange diamond) defines the date and depth of the central dip feature.

The observation on JD 2457974 was made by Dr Boyajian using the Las Cumbres Observatory, Tiede Observatory in Tenerife, Spain (LCO - TFN), in r'-band (from Dip Update 53n, WTF blog http://www.wheretheflux.com/single-post/2017/08/09/Dip-update-53n ).

In order to assess agreement of HAO measurements with the SPOT model solution for Kepler data, the HAO observations in August 2017 were averaged to form 21 binned (hourly) groups. We emphasize that the SPOT model was derived (in 2016) as a fit to *Kepler* data; it was not adjusted to provide a fit to the HAO 2017 August observations. The RMS between the August HAO observations and the SPOT model is ~ 0.1% (~ 1 mmag); 17 of the 21 observations exhibit a variance with respect to the model of < 0.1% (< 1 mmag). The ingress/egress gradients between nights, and calculated for within an observing session with two or more hours of observations, were also in agreement with gradients predicted by the SPOT model.

| Ingress JD | Variance with SPOT model | Egress JD | Variance with SPOT model |
|---|---|---|---|
| 57967.7 | -0.1% | | |
| 57968.8 | 0.2% | 57976.7 | 0.0% |
| 57968.9 | 0.1% | 57976.8 | 0.0% |
| 57969.7 | 0.1% | 57976.9 | 0.0% |
| 57969.7 | 0.2% | 57978.7 | 0.0% |
| 57970.7 | -0.1% | 57978.7 | 0.1% |
| 57971.7 | -0.2% | 57979.7 | 0.0% |
| 57972.7 | 0.0% | 57979.7 | 0.0% |
| 57972.8 | -0.1% | 57980.7 | -0.1% |
| 57973.7 | -0.2% | 57981.6 | 0.0% |
| 57973.7 | -0.1% | 57981.7 | 0.0% |

**Table 5.1:** Variance of HAO August 2017 observations and SPOT model of the D1540 transit

Considering the excellent agreement between the 21 HAO observations and the SPOT model fitted to the Kepler D1540 dip of 1601 days earlier, combined with similar shapes for ingress and egress, and in addition the sharp central dip, it is highly suggestive that the August dimming event was a repeat of the D1540 dip.

Indirect support for the suggestion that something is in a ~ 1600-day orbit also comes from a study by Ballesteros et al., (2017) showing that weekly variability of *Kepler* data exhibits a symmetrical pattern about ~ D790, where a minimum of variability occurs. The maximum of variability would correspond to ~ D1600, which corresponds to our suggested periapsis; the D790 minimum corresponds to our suggested apoapsis.

## 6. CONSTRAINTS ON THE ORBIT

The SPOT Model of the D1540 transit calls for an outer ring approximately 0.1 AU in radius and a transit velocity of ~ 28 km s$^{-1}$. These parameters, along with the proposed 1601-day orbital period, provide some constraints on the eccentricity and mass of the object the ring material orbits.

The circular orbital velocity of an object in a 1601-day orbit (3.02 AU) around a 1.43 $M_S$ star is ~ 20.5 km s$^{-1}$. If the orbit were eccentric, with a periapsis of 2.0 AU, and if the periapsis location was close to the line-of-sight to the star, the transit velocity would be 28 km s$^{-1}$.

The SPOT model suggests that the ring system transit velocity increases during the transit, which would occur if periapsis is not aligned close to our line-of-sight to the star, and that the maximum orbital velocity may be significantly greater than 28 km s$^{-1}$; this implies that periapsis is likely to be closer to 1.5 AU than 3.0 AU, and this would require an eccentricity ~ 0.5.

The Hill sphere at 1.5 AU for a stable 0.1 AU radius prograde ring system requires a mass for the D1540/August 2017 object of at least 14 $M_J$, which is in the brown dwarf region.

Based on the interpretation of the August 2017 transit as a repeat of D1540, and assuming periapsis occurs near the end of the 1-year fade, a brightening should take place when the proposed brown dwarf passes periapsis. October to December 2017 would be the dates for the brightening to occur. The observed brightening in October is compatible with such an orbit.

The proposed ring system around the brown dwarf may exhibit equinox transitions (front-lit to back-lit) like the rings of Saturn. If the brown dwarf is in an eccentric orbit with a configuration similar to that suggested in Fig 6.1, there should be asymmetry in the yearlong U-shaped fade event marked by a rapid brightening after periapsis compared to the slower dimming phase.

One feature of the proposed configuration is that a set of dips should occur during a few months for each 1601-day recurrence of the proposed brown dwarf transit. Hippke and Angerhausen (2017) questioned the notion that long term flux variations are cyclical due to the lack of ASAS data at the back-projected dates in 2008. We suggest that the sparse ASAS measurements in 2008 (as well as the unclear SuperWASP calibration accuracy between seasons, which don't show a fade) do not exclude the presence of a fade at that time.

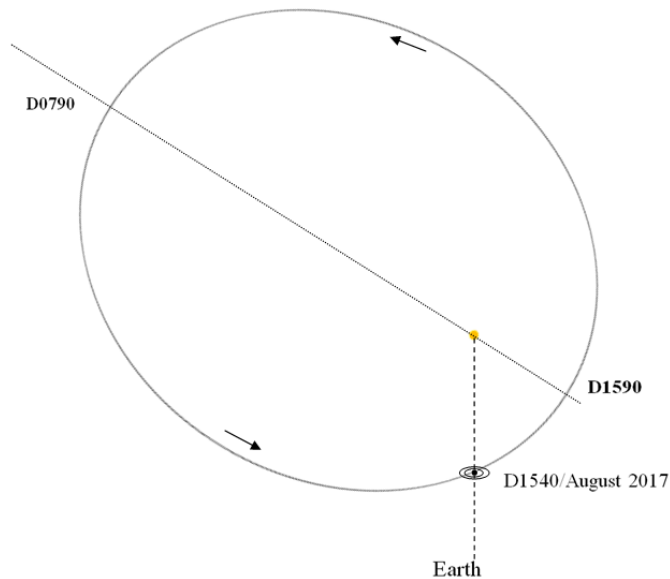

**Figure 6.1:** System Configuration of the 1601-day elliptical orbit of the proposed brown dwarf and ring system (not to scale).

The authors are preparing another paper that will contain a detailed discussion of the system configuration, including possible satellites of the proposed brown dwarf. This includes modeling of D792 and deconstruction of the D1519 and D1568 dips into individual transits of comas and dust clouds that could be produced by sublimation of icy moons.

## 7. DISCUSSION AND CONCLUSIONS

We have identified three characteristics of KIC8462 photometric behavior that support the assertion of a repeating pattern with a ~ 1600-day interval: 1) a yearlong fade of 1 or 2 % has appeared in the Kepler data as well as recent ground-based data 1600 days later (the U-shaped fade), 2) most dips occur during the last half of both yearlong U-shaped fades, and 3) the Kepler fade D1540 has a counterpart with a remarkably similar structure that appeared 1601-days later.

We used a 2D transit model consisting of a ring system surrounding an object with mass and size corresponding to a brown dwarf for fitting the D1540 data. These comparisons show that the same model, without modification, fits the HAO 2017 August dip with a precision of ~ 0.1 %. We have not attempted to explain how such a ring system could come into existence because our intent is limited to showing that a natural system is capable of accounting for observed brightness variations without having to invoke the more complicated speculation of alien mega-structures.

There is a clear fade event in the last year of the *Kepler* data (MS16), and most dips occur during the last six months of this fade. A similar fade event is identified in the HAO observations: a year-long U-shaped fade feature, with a series of dips occurring in the last six months. The resemblance of these two features is striking.

The August 2017 and D1540 dips are interpreted as possibly due to transits of a brown dwarf and associated ring system in a 1601-day elliptical orbit. If a brown dwarf in this configuration is confirmed by future radial velocity measurements, it could be speculated that the orbit has been disrupted from beyond the 'snow line' by Lidov-Kozai oscillations as describe by Metzger et al. (2017).

We look forward to future work evaluating the feasibility that transiting icy moons of the proposed brown dwarf, sublimating near periapsis like comets, could be sufficiently active in producing dust that extends to large enough projected area so as to provide an explanation for the significant dips observed by *Kepler* and the recent May to October 2017 dips observed from the ground. The dust responsible for these dips may contribute to the long term variation in flux, described by Simon at al. (2017). The proposed brown dwarf and extensive ring system, ~ 0.2 AU across, could also contribute to longer term flux variations as it orbits the star.

Based on this proposed 1601-day period we make the following predictions:

1) a rapid brightening in flux by about 1-2% should occur during October to December 2017 (this is compatible with the brightening observed during October),
2) a repeat of D1540 should occur on 27 December 2021,
3) another U-shaped fade should occur in 2021, and
4) a set of dimming events, similar to those in May-October 2017, should occur between October 2021 and January 2022.

The results reported in this paper are merely an early step in understanding the KIC8462 system because we have intentionally limited our analysis to the geometry of one dust extinction structure (a ring system) whose transit could account for a fade feature in the *Kepler* data, D1540, which we suggest was observed 1601-days later by ground-based observatories. We have not considered how such a ring structure could be created and maintained for at least 4.4 years, nor have we considered the physical processes that could produce the ring system dust with the necessary extinction properties. Considerable ground-work on these matters has been reported by Neslusan and Budaj (2016), and we look forward to future investigations that combine our updates to transit geometry and orbit period with physical mechanisms.

This paper demonstrates the utility of high precision amateur photometric observations and of light curve simulation models in understanding complex shaped transit photometry.

**ACKNOWLEDGEMENTS**

We thank Dr. T. Boyajian for making her LCO observations available in the public domain, which allowed us to learn that the August 2017 dimming event exhibited a brief and deep central dip on JD 2457974. Author BLG wants to express appreciation for Dr. T. Boyajian for encouraging him to combine HAO unfiltered observations in 2015 with V-band observations in 2016 for evaluating long-term fade structure.

We also thank OHIO State University and the Gordan and Betty Moore Foundation for making the ASAS SN data publicly available.

# Appendix

SPOT Simulation Video D1540
https://youtu.be/4fQfaMxc9lE

# References

Aizawa M., et al., 2017, *AJ*, 153, 193

Ballesteros, F. J., et al., 2017, preprint *(arXiv:1705.08427)*

Boyajian T. S., et al., 2016, *MNRAS*, 457, 3988

Chen J., and Kipping D. M., 2016, *ApJ* 834, 17

Hippke M., and Angerhausen D., 2017, preprint, (arXiv:1710.05000)

Kenworthy, M. A., & Mamajek, E. E., 2015, *ApJ*, 800, 126

Meng H. Y. A., et al., 2017, preprint, (arXiv: 1708.07556)

Metzger B. D., Shen K. J., and Stone N., 2017, *MNRAS*, 468, 4399

Montet B. T., and Simon J. D., 2016, *ApJ*, 830, L39

Neslusan, L. and J. & Budaj, 2017, *A&A*, 600, A86 (arXiv:1612.06121v2)

Rieder S., and Kenworthy M. A., 2016, *A&A*, 596, A9

Sacco G., Ngo L., and Modolo J., 2017, preprint, (arXiv:1710.01081)

Schlichting H. E., and Chang P., 2011, *ApJ*, 734, 117

Simon J. D., et al., 2017, preprint, (arXiv:1708.07822)

Sing D. K., 2009, *A&A*, 510, 21